\documentclass{article}
\usepackage[T1]{fontenc}
\usepackage{setspace}
\IfFileExists{url.sty}{\usepackage{url}}
                      {\newcommand{\url}{\texttt}}

\makeatletter

\providecommand{\LyX}{L\kern-.1667em\lower.25em\hbox{Y}\kern-.125emX\@}

 \newenvironment{lyxlist}[1]
   {\begin{list}{}
     {\settowidth{\labelwidth}{#1}
      \setlength{\leftmargin}{\labelwidth}
      \addtolength{\leftmargin}{\labelsep}
      }}
   {\end{list}}
 \newcommand{\lyxaddress}[1]{
   \par {\raggedright #1 
   \vspace{1.4em}
   \noindent\par}
 }

\usepackage[T1]{fontenc}

\makeatletter

\date{}
\makeatother
\makeatother

\begin{document}

\title{Tables of Nuclear Cross Sections and Reaction Rates: an Addendum to 
the Paper
``Astrophysical Reaction Rates from Statistical Model Calculations''}

\author{Thomas Rauscher and Friedrich-Karl Thielemann}

\maketitle

\lyxaddress{Departement f\"ur Physik und Astronomie, Universit\"at Basel, 
Klingelbergstr.\ 82, CH-4056 Basel, Schweiz (Switzerland)}

\begin{abstract}
\noindent In a previous publication {[}Atomic Data and Nuclear Data Tables 75,
1 (2000){]}, we had given seven parameter analytical fits to theoretical
reaction rates derived from nuclear cross sections calculated in the 
statistical model (Hauser-Feshbach formalism) for targets with 
\( 10\leq Z\leq 83 \) (Ne to Bi)
and for a mass range reaching the neutron and proton driplines. Reactions 
considered were (n,$\gamma$), (n,p), (n,$\alpha$), (p,$\gamma$), (p,$\alpha$), 
($\alpha$,$\gamma$), and their inverse reactions.
Here, we present the theoretical nuclear cross sections and astrophysical 
reaction rates from which those rate fits were derived, and we provide 
these data as on-line electronic files. Corresponding to the fitted rates, 
two complete data sets are provided, one of which includes a phenomenological 
treatment of shell quenching for neutron-rich nuclei.
\end{abstract}
\tableofcontents{}

\section{Introduction}

In a previous paper by Rauscher and Thielemann (\cite{RT00}, hereafter referred
to as RATH) we presented fits to nuclear reaction rates covering a wide range 
of isotopes and most of the nuclear chart.
The data sets of reaction rate fits in that paper were mainly suited to direct 
application in astrophysical simulations.
However, in certain computational approaches it is preferable to have the rates 
in the original tabular format
without any further fitting procedure which, on one hand, may be advantageous
for the speedy calculation of a rate at a given temperature with an economical
use of computer memory but, on the other hand, may introduce additional 
inaccuracies.
Furthermore, for nuclear physics applications it is often preferable to have
the tables of nuclear cross sections directly.
Therefore we thought it
worthwhile to also publish the original cross sections and reaction rates
from which the fits given in RATH have been derived; we provide these 
extensive data in the form of electronic tables. 
(It should be noted that ``data'' in
this context refers to results of theoretical calculations; wherever there
is a reference to experimental data, it is stated explicitly.)

As in RATH, we provide here two sets of calculations with input from two 
different mass models, the finite range droplet model (FRDM) \cite{frdm} and 
an extended Thomas-Fermi approach with Strutinski integral (ETFSI-Q) 
\cite{etfsiq}. Experimental masses and level properties had been used in
the calculations where available. In
addition to cross sections and reaction rates, particle separation energies
are given. In order to facilitate a comparison to other calculations, all the
experimental level information used in our calculations is shown in a separate
table.

Here, we do not want to repeat the detailed definitions of the statistical model
and of reaction rates already given in RATH. Therefore we limit ourselves to
the most concise form and refer the reader to RATH for more
details. In the following, we concentrate on the explanation of how to extract
values from the on-line tables and on giving a few \emph{caveats} when using
the tabulated information .

\section{Tabulated information}

\label{sec:tabinfo}The data set consists of seven tables provided as 
separate on-line files.
The four files described in Secs.~\ref{cs_exo} and \ref{astro_exo} contain
the raw data used to generate the fits given in RATH. Further information which
may be useful for comparison to experiments or other calculations is given in
three additional files (see Secs.~\ref{cs_endo} and \ref{levels}). 
All files are formatted in such a way as to be machine
readable without losing the accessibility to the human reader. That
way, the additional information provided here can best serve the needs of 
several
specialized fields of nuclear physics as well as astrophysics. Definitions of
cross sections and reaction rates as well as further details of the statistical
model calculations are already given in RATH.

\subsection{Laboratory Cross Sections For Capture Reactions and Exothermic 
Reactions}

\label{cs_exo}Reaction cross sections involving neutrons, protons and 
\( \alpha  \) particles
as projectiles or ejectiles have been calculated for targets from Ne up to,
and including, Bi. The results of two different sets of calculations are given
here in two files. They differ in the mass model used, which enters into the
computation of the separation energies and \( Q \) values as well as into the
microscopic input to the level density calculation. One set was calculated 
employing
the well-known FRDM mass model \cite{frdm}, the other set employed the ETFSI-Q
model \cite{etfsiq} which includes a phenomenological approximation of 
shell-quenching effects (see RATH for further details on the mass models).

Tabulated are the laboratory cross sections for capture reactions and for 
exoergic
particle reactions. These assume that the target is in the ground state and
cannot, in general, be used for computation of an astrophysical rate as given
in RATH (cf.~Eqs. (1), (2), (10) in RATH). However, they can be directly 
compared
to experimentally measured data. The isotope range is as specified
in Table A of RATH, which is repeated here again for convenience. Cross 
sections for exothermic (\( Q\geq 0 \)) reactions with particles in the 
exit channel are given here as well as all capture reactions
regardless of the \( Q \) value. All those reactions are contained in one file
each for each mass model, sorted by target (charge and mass number) and 
projectile
(neutron, proton, \( \alpha  \) particle), in that order. The energy grid is
different for each combination of target and projectile as it was chosen in
such a way as to provide the best grid for the numerical integration involved in
the computation of the astrophysical reaction rate. This means that, e.g., the
energies are densely clustered around channel openings and more widely spaced
in regions of ``well-behaved'' cross sections. The energies are given in MeV;
the cross sections are given in barn 
(1 barn = 10\( ^{-28} \) m\( ^{2} \) = 100 fm\( ^{2} \)).

\begin{table}
\small
\begin{center}
TABLE A\\
Isotope Range of the On-line Cross Section and Reaction Rate Files

\vspace{0.2cm}

\begin{tabular}{rrrrrrrrrrr}
\hline\\
 & \multicolumn{2}{c}{FRDM} & \multicolumn{2}{c}{ETFSI-Q}& & &
\multicolumn{2}{c}{FRDM} & \multicolumn{2}{c}{ETFSI-Q} \\
&\multicolumn{2}{c}{\hrulefill}&\multicolumn{2}{c}{\hrulefill}& & &
\multicolumn{2}{c}{\hrulefill} & \multicolumn{2}{c}{\hrulefill}\\
$Z$ & $N_{\rm min}$ & $N_{\rm max}$ &$N_{\rm min}$ & $N_{\rm max}$ &
&
$Z$ & $N_{\rm min}$ & $N_{\rm max}$ &$N_{\rm min}$ & $N_{\rm max}$  \\
\hline\\
  8 &   5 &  10 &     &     &&  47 &  41 & 113 &  41 & 111\\
  9 &   5 &  28 &     &     &&  48 &  42 & 115 &  42 & 112\\
 10 &   5 &  31 &     &     &&  49 &  43 & 117 &  43 & 113\\
 11 &   6 &  33 &     &     &&  50 &  44 & 119 &  44 & 114\\
 12 &   7 &  35 &     &     &&  51 &  46 & 121 &  46 & 115\\
 13 &   8 &  38 &     &     &&  52 &  47 & 124 &  47 & 124\\
 14 &   8 &  40 &     &     &&  53 &  48 & 126 &  48 & 126\\
 15 &   8 &  42 &     &     &&  54 &  49 & 128 &  49 & 128\\
 16 &   8 &  44 &     &     &&  55 &  51 & 130 &  51 & 130\\
 17 &   9 &  46 &     &     &&  56 &  52 & 133 &  52 & 132\\
 18 &   9 &  49 &     &     &&  57 &  53 & 135 &  53 & 133\\
 19 &  10 &  51 &     &     &&  58 &  55 & 137 &  55 & 134\\
 20 &  10 &  53 &     &     &&  59 &  56 & 139 &  56 & 135\\
 21 &  11 &  55 &     &     &&  60 &  58 & 141 &  58 & 136\\
 22 &  12 &  58 &     &     &&  61 &  59 & 144 &  59 & 137\\
 23 &  13 &  60 &     &     &&  62 &  61 & 146 &  61 & 138\\
 24 &  14 &  62 &  18 &  62 &&  63 &  62 & 148 &  62 & 139\\
 25 &  15 &  64 &  18 &  64 &&  64 &  64 & 150 &  64 & 150\\
 26 &  16 &  66 &  19 &  66 &&  65 &  65 & 153 &  65 & 152\\
 27 &  17 &  69 &  19 &  67 &&  66 &  67 & 155 &  67 & 154\\
 28 &  18 &  71 &  20 &  68 &&  67 &  69 & 157 &  69 & 155\\
 29 &  19 &  73 &  21 &  69 &&  68 &  70 & 159 &  70 & 156\\
 30 &  21 &  75 &  22 &  70 &&  69 &  72 & 161 &  72 & 157\\
 31 &  22 &  77 &  23 &  71 &&  70 &  73 & 164 &  73 & 158\\
 32 &  23 &  80 &  24 &  72 &&  71 &  75 & 166 &  75 & 159\\
 33 &  24 &  82 &  25 &  73 &&  72 &  77 & 168 &  77 & 160\\
 34 &  25 &  84 &  26 &  84 &&  73 &  78 & 170 &  78 & 161\\
 35 &  26 &  86 &  27 &  86 &&  74 &  80 & 173 &  80 & 162\\
 36 &  27 &  88 &  28 &  88 &&  75 &  81 & 175 &  81 & 163\\
 37 &  29 &  91 &  29 &  89 &&  76 &  83 & 177 &  83 & 177\\
 38 &  30 &  93 &  30 &  90 &&  77 &  85 & 179 &  85 & 179\\
 39 &  31 &  95 &  31 &  91 &&  78 &  87 & 182 &  87 & 182\\
 40 &  32 &  97 &  32 &  97 &&  79 &  88 & 184 &  88 & 184\\
 41 &  33 &  99 &  33 &  99 &&  80 &  90 & 186 &  90 & 186\\
 42 &  35 & 102 &  35 & 102 &&  81 &  92 & 188 &  92 & 188\\
 43 &  36 & 104 &  36 & 104 &&  82 &  93 & 191 &  93 & 191\\
 44 &  37 & 106 &  37 & 106 &&  83 &  95 & 193 &  95 & 193\\
 45 &  38 & 108 &  38 & 108 &&  84 &  98 & 193 &  98 & 193\\
 46 &  40 & 110 &  40 & 110 &&  85 & 101 & 195 & 101 & 195\\
\hline
\end{tabular}
\end{center}

\vspace{0.1cm}

\begin{footnotesize}
\emph{Note.} Given are the charge number \protect\( Z\protect \)
and the lower and upper limits \protect\( N_{\mathrm{min}}\protect \)
and \protect\( N_{\mathrm{max}}\protect \) of the neutron number in the
isotopic chain.
The complete files \textit{cs\_frdm.asc, cs\_etfsiq.asc, 
cs\_frdm\_endo.asc, cs\_etfsiq\_endo.asc, rates\_frdm.asc, rates\_etfsiq.asc},
and \textit{levels.asc} are available electronically.
\end{footnotesize}
\end{table}
Provided as additional information are the particle separation energies 
in the compound nucleus, from
which the reaction \( Q \) values given in RATH were derived.

\subsubsection{Instructions for the Use of the Cross Section Files}

\label{cs_exo_ex}The file can either be browsed by using a text editing program
or a code specifically written for reading the provided data. Table I shows
an example of a segment from the FRDM file. The data blocks are sorted first by 
target nucleus (charge, mass number)
and then by projectile. For each target-projectile combination, the capture 
cross sections are listed for all calculated projectile energies
regardless of $Q$ value. Additionally, cross sections
are given for those reactions with particles in the exit channel, having 
\( Q\geq 0 \). Thus, each data block
will have a different number of lines, depending on the number of calculated
energies and exoergic reaction channels. The lines within each block
are organized as follows.
Each data block for a specific combination of target and projectile starts 
with an
identifying line specifying, in that order: charge and mass number of the 
target;
a digit specifying the projectile (0 for neutron, 1 for proton, 2 for $\alpha$
particle); the number \( L \) of calculated energies; indices \( b \) and
\( c \) indicating whether or not a reaction with particles in the exit
channel has $Q\geq 0$ and is therefore listed ($=1$) or not listed ($=0$);
charge $Zc$ and mass number $Ac$ of the compound nucleus; and the particle 
separation energies \( S_{n} \), \( S_{p} \), \( S_{\alpha } \) of the 
compound nucleus.
The end of the line gives again target isotope ($T$, projectile ($x$)
and the three possible exit
channels denoted by roman letters (g=$\gamma$, n, p, and a=$\alpha$). 
Following the first line are the \( L \) center-of-mass
energies in MeV at which the cross sections were calculated. There are at most
eight entries per line. Then, \( L \) capture cross sections are printed,
regardless of the \( Q \) value. Values for the other reaction channels are 
only given if \( Q\geq 0 \). If \( b=0 \) and/or $c=0$, then no cross section 
values are printed for the respective exit channel 1, 2. Thus, the capture 
cross sections are followed by \( L\times (b+c) \) cross sections and therefore 
a total of \( L\times (1+b+c) \)
cross sections is given. A new data block starts with the next identification 
line after the last line of cross sections.

As an example we consider the reaction \( ^{70} \)Ga(p,n)\( ^{70} \)Ge. Since
\( Q=S_{p}( \)\( ^{71} \)Ge\( )-S_{n}( \)\( ^{71} \)Ge\( )>0 \) the cross
sections can be found in the regular cross section file. Table I shows a small
part of that table. Looking for \( ^{70} \)Ga, e.g.\ by a text search for 
``ga70'',
one first encounters the data block for \( ^{70} \)Ga+n. Farther down one finds
the identification line for \( ^{70} \)Ga+p since for the same target isotope 
the data blocks are sorted
by reaction. With a computer code it would be more economical to search for the
numeric codes for charge, mass number and projectile at the beginning of the
line. The relevant identification line tells us that the cross sections were 
calculated at 61 energies and that both the values for the (p,n) as well as 
the (p,\( \alpha  \))
reaction are printed. This line is followed by the 61 center-of-mass energies,
then 61 capture cross sections, 61 (p,n) cross sections and, finally, 61 
(p,\( \alpha  \))
cross sections. Thus, for instance, the value of the 
\( ^{70} \)Ga(p,n)\( ^{70} \)Ge
cross section at 2.3022 MeV is given as 2.964 millibarn. The RATH tables should
always be consulted to determine if the statistical model is suited for 
predicting cross sections for a specific reaction at a given rate 
(see Sec.~\ref{sec:appl}).

The values for the inverse reaction \( ^{70} \)Ge(n,p)\( ^{70} \)Ga cannot
be extracted from the same file as \( Q<0 \) for this reaction. An inspection
of the identification line for \( ^{70} \)Ge+n farther down shows that 
\( b=0 \) and that therefore no cross sections are listed for this reaction. 
Particle reactions with \( Q<0 \) are given in a separate file, see 
Sec.~\ref{cs_endo}.

The above instructions are relevant to the posted cross section files
\textit{cs\_frdm.asc}, \textit{cs\_etfsiq.asc}, \textit{cs\_frdm\_endo.asc}, and
\textit{cs\_etfsiq\_endo.asc}.

\subsection{Astrophysical Reaction Rates For Capture Reactions and Exothermic 
Reactions}

\label{astro_exo}Reaction rates are given in two separate files, one for
each mass model, for the reactions as included in the cross section
listings. The rates were computed on a grid of 24 temperatures:
\( T_{9}= \)0.1, 0.15, 0.2, 0.3, 0.4, 0.5, 0.6, 0.7, 0.8, 0.9, 1.0, 1.5, 2.0,
2.5, 3.0, 3.5, 4.0, 4.5, 5.0, 6.0, 7.0, 8.0, 9.0, 10.0 (\( T_{9} \) is given
in 10\( ^{9} \) K). The reactions are sorted as before. For each reaction,
the laboratory rate \( N_{A}\left\langle \sigma v\right\rangle 
^{\mathrm{lab}} \)
as well as the stellar rate \( N_{A}\left\langle \sigma v\right\rangle ^{*} \)
is given. The laboratory rate was computed directly from the cross sections
discussed in Sec.~\ref{cs_exo} and can be used for comparison with rates derived
from experiment. Only the stellar rate considers the thermal excitation in a
stellar plasma and should be used exclusively in astrophysical calculations.
It was computed making use of the stellar transmission functions (see Eqs. (2)
and (3) in RATH). The stellar reaction rates were the basis for the RATH fits.
The rates are given in units of cm\( ^{3} \) s\( ^{-1} \) mol\( ^{-1} \).

\subsubsection{Instructions for the Use of the Rate Files}

\label{astro_exo_ex}An example for a segment of the FRDM rate file
is given in Table II.
The structure of the table is somewhat similar to the cross section table as
described in Sec.~\ref{cs_exo_ex}. The data blocks are sorted first by target 
nucleus (charge, mass number) and then by projectile. For each target and 
projectile, the capture rates are printed, regardless of $Q$ value. The rates 
are also printed for those particle
exit channels for which \( Q\geq 0 \). Thus, each data block will have a 
different number of total lines, depending on the number of exoergic particle 
channels.
A data block starts with an identification line giving charge and mass number of
target, the projectile (0 for neutron, 1 for proton, 2 for $\alpha$
particle), and indices \( b \) and
\( c \) specifying whether a particle exit channel is listed. 
The digits \( b \) and \( c \) can only assume the values
0 (not listed) or 1 (listed). The end of the line gives again target isotope, 
projectile and the three possible exit
channels. This is followed by 24
laboratory rates (three lines with eight entries each) calculated at the 
temperatures given in the previous section. A blank line separates the last 
line of the laboratory rates from the following three lines with stellar rates.
Laboratory and stellar rates are given for the capture reaction regardless of 
the \( Q \) value. This is followed by the laboratory and stellar rates of the 
particle exit channels for which \( b=1 \) and/or $c=1$, i.e.\ with 
\( Q\geq 0 \). Thus, there are \( 2*24*(1+b+c) \) rates printed in total.

As an example, the reaction \( ^{70} \)Ga(p,n)\( ^{70} \)Ge is again considered.
As before in Sec.~\ref{cs_exo_ex}, since 
\( Q=S_{p}( \)\( ^{71} \)Ge\( )-S_{n}( \)\( ^{71} \)Ge\( )>0 \)
the reaction rates can be found in the regular rate file. Table II shows a small
part of that table. Looking for \( ^{70} \)Ga, for instance by a text search 
for ``ga70'', one first encounters the data block for \( ^{70} \)Ga+n. 
Farther down one finds
the data block \( ^{70} \)Ga+p since for the same target isotope the 
data blocks are sorted
by reaction. With a computer code it may be more convenient to search for the
numbers for charge, mass number and projectile at the beginning of the line.
The relevant identification line tells us that the values for the (p,n) as
well as the (p,\( \alpha  \)) reaction are printed. Unlike
the cross section table, which starts with $L$ projectile energy values,
the temperature grid is not given here because the rates are always calculated
at the same 24 stellar temperatures given above. The first line is 
immediately followed by the rates for the capture reaction, first 24 values 
of \textit{laboratory} rates, then 24 \textit{stellar} rates. These are 
followed by pairs of 24 laboratory
and stellar rates each for the (p,n) and the (p,\( \alpha  \)) reactions. Thus,
for instance, the value of the \( ^{70} \)Ga(p,n)\( ^{70} \)Ge laboratory
rate at \( T_{9}=9 \) (next to last position in the third line)
is given as 7.27\( \times  \)10\( ^{6} \) cm\( ^{3} \)
s\( ^{-1} \) mol\( ^{-1} \). This should be compared to the stellar rate
of 6.94\( \times  \)10\( ^{6} \) cm\( ^{3} \) s\( ^{-1} \) mol\( ^{-1} \).
The RATH tables should always be consulted to determine if the statistical model
is suited for predicting cross sections for a specific reaction at a given rate
(see Sec.~\ref{sec:appl}).

The values for the inverse reaction \( ^{70} \)Ge(n,p)\( ^{70} \)Ga cannot
be extracted from the file as \( Q<0 \) for this reaction. An inspection of
the identification line for \( ^{70} \)Ge+n further down shows that 
\( b=0 \)
and that therefore no rates are printed for this reaction. Reverse stellar
rates for particle reactions can easily be computed as explained in the 
following section.

The above instructions are relevant to the posted rate files
\textit{rates\_frdm.asc} and \textit{rates\_etfsiq.asc}.

\subsection{Reverse Stellar Rates and Endothermic Laboratory Cross Sections}

\label{cs_endo}The data presented in Secs.~\ref{cs_exo} and \ref{astro_exo}
correspond to the RATH tables. Since reverse astrophysical rates can easily
be calculated by detailed balance, the previous RATH work only gave such rates
explicitly for exothermic reactions and capture reactions and provided the means
to derive the fit parameters for the reverse rates. The reverse stellar
rate is also
not explicitly given here but can easily be derived for each temperature with
the following formulas. For reactions with particles in all channels use
\begin{equation}
N_{A}\left\langle \sigma v\right\rangle ^{*}_{\mathrm{rev}}=N_{A}\left\langle 
\sigma v\right\rangle ^{*}\frac{G_{i}(T)}{G_{m}(T)}e^{f}\quad .
\end{equation}
In order to compute a photodisintegration rate \( \lambda _{\gamma }^{*} \)
(in s\( ^{-1} \)) from a capture rate use
\begin{equation}
\lambda ^{*}_{\gamma }=\frac{G_{i}(T)}{G_{m}(T)}T_{9}^{3/2}N_{A}\left\langle 
\sigma v\right\rangle ^{*}e^{f}\quad .
\end{equation}
 The exponent \( f \) is defined as
\begin{equation}
f=a_{0}^{\mathrm{rev}}-a_{0}-\frac{11.6045Q}{T_{9}}\quad .
\end{equation}
The quantities \( G_{i} \), \( G_{m} \) are the partition functions of the
target and residual nucleus, respectively, in the forward reaction, i.e.\ the
reaction \( x+i\rightarrow m+y \), where \( x \) can be a particle and \( y \)
can be a particle or a \( \gamma  \)-ray (compare Eqs.~12--14 in RATH). The
parameters \( a_{0} \), \( a_{0}^{\mathrm{rev}} \) and the reaction \( Q \)
value (in MeV) for the forward reaction
as well as the required partition functions can be looked up
in the RATH tables.  The \( Q \) value can also
be calculated from the particle separation energies given with the cross sections.

It has to be emphasized that detailed balance can only be applied to stellar
rates and not to laboratory rates or cross sections with the target being in
the ground state. When comparing to experiments it can sometimes be useful to
also have access to cross sections for reactions with negative \( Q \) values.
For that purpose, calculated cross sections for such endothermic
reactions are given in two additional files:
\textit{rates\_frdm\_endo.asc} and \textit{rates\_etfsiq\_endo.asc} 
(one file for each mass model). The format is the same
as for the cross sections discussed in Sec.~\ref{cs_exo}. Here, only the 
energies are given for which the cross sections are not zero, i.e.\ above the 
particle thresholds. Cross sections at lower energies can be assumed to be zero.

Photodisintegration cross sections for targets in the ground state are not given
here. They are discussed elsewhere \cite{moh00,vog00} and a table will be 
published separately \cite{rau01}.

\subsection{Experimental Nuclear Level Schemes}

\label{levels}The experimental level information utilized in the statistical
model calculations is provided in the last file. Up to 20 experimental levels 
(including the ground state) were considered. The states are taken from 
\cite{toi96},
up to the first level for which the spin or parity assignment was not known.
Above this cut-off a theoretical nuclear level density \cite{RTK97}
was employed. Ground state spin and parities are known for many unstable nuclei.
Far off stability, where experimental values are not known, ground state spins
and parities are derived from the neutron and proton single-particle spins given
in \cite{mnk97} by applying Nordheim rules. However, tabulated here are only
the experimental spin and parity values used because the complete list
of ground state spins is already given in RATH. This information is important 
when comparing with other statistical model calculations.

\subsubsection{Instructions for the Use of the Nuclear Level File}

\label{levels_ex}An example is shown in Table III. The isotopes are ordered
first by charge and then by mass number. The data block for each isotope starts 
with an identification line specifying, in this order: charge and mass number 
of the isotope, number \( L \) of experimental nuclear levels given, and finally
the isotope again written in element notation. The succeeding lines specify the 
\( L \) nuclear levels, giving for each
level the excitation energy (in MeV) as well as the parity and the spin 
assignment.
The parity is coded as a sign before the spin. There are at most five levels per
line. A new isotope entry again starts with an identification line.

As an example, consider the nucleus \( ^{70} \)Ga in Table III, which in
turn shows a small fraction of the full file. It can be seen that five
experimental levels were used in the calculation, including the ground state. 
The list starts with the $J^{\pi}=1^+$
ground state, while the third excited state of \( ^{70} \)Ga is located
at an energy of 0.6909 MeV and has a spin and parity assignment of 
\( J^{\pi }=2^{-} \).

The above instructions are relevant to the posted file
\textit{levels.asc}.

\subsection{Caveats}

\label{caveats}The cross sections and reaction rates provided here should only
be used keeping in mind the fundamental limitations of the model as
discussed below.

\subsubsection{Applicability of the Statistical Model}

\label{sec:appl}The statistical model can be applied provided that the use
of averaged transmission coefficients (Eq.~3 in RATH) is permitted. This will
be the case for high level densities with completely overlapping resonances,
typical for the compound nucleus reaction mechanism. For light nuclei, 
decreasing particle separation energies, or at shell closures, level densities 
will eventually
become too low for the application of the statistical model at astrophysical
temperatures. In those cases, single resonances and contributions from the 
direct
reaction mechanism have to be taken into account \cite{raubi98}. Based on a
level density description, a quantitative criterion for the applicability was
derived recently~\cite{RTK97}. The lower temperature limit of the validity
of the statistical model for the calculation of reaction rates is already given
in the RATH tables. \textit{It is advisable to check those limits before
using the cross sections and rates given in this paper.}

In principle, a similar criterion can also be derived for cross sections when
comparing to experimental data. In the case of a nuclear experiment, the energy
distribution over which one has to average is not determined by a 
Maxwell-Boltzmann
velocity distribution but by the uncertainty in the beam energy and the energy
straggling in the target. Since these differ depending on the experimental setup
it is not possible to give a global criterion. However, as a first approximation
one can use the temperature limits given in RATH.

\subsubsection{Validity of the Predictions}

\label{valid}It should be noted that only purely theoretical rates are given
here which do not use any direct experimental information (except for nuclear
masses and ground and excited state information where available). The methods 
to predict nuclear properties needed in the statistical model calculations are 
chosen to be as reliable as possible in order to retain predictive power. 
Nevertheless, the uncertainties and some known weaknesses (e.g.\ of the FRDM 
around shell closures)
of the input will be reflected in the predicted values. This is a compromise
which may lead to locally enhanced inaccuracies but it emphasizes the importance
of reliable predictions of rates far off stability (see RATH). If one would
use experimental data (such as level densities) and locally tuned 
parametrizations
of nuclear properties (optical potentials, Giant Dipole Resonance widths) as
input for our statistical model calculation, a further increase in accuracy
could be achieved. Since the main scope of this work is the prediction of 
nuclear reaction rates for astrophysics for experimentally unaccessible nuclei,
a \emph{global} approach is better suited. Therefore, in comparison to 
experimental work deviations within a factor of \( 1.5-2 \) could appear 
although the average deviation will be smaller. For neutron capture it has been 
shown that the average deviation is about a factor of \( 1.3-1.4 \) 
\cite{RTK97}.

\section{Summary}

\label{sec:summary}Nuclear cross sections and thermonuclear reaction rates
for neutron-, proton- and \( \alpha  \)-induced reactions and their inverses
have been calculated in the statistical model and published in the previous
paper RATH \cite{RT00}. All cross sections and rates from Ne to Bi are given 
from proton dripline to neutron dripline, thus covering a considerable part of 
the nuclear chart. Fits to reaction rates have been provided in RATH
for two sets of rates, calculated
with input from two different mass models. Here, we make accessible the input
data for the original calculation, containing nuclear cross sections as well as
reaction rates. Excitation energy, spin and parity are quoted for the 
experimental
level information used in the calculations to facilitate comparison with other
models. Further details on the underlying models are given in RATH.

In real applications, these rates should be supplemented or replaced by 
experimental rates as they become available. Such a combination of theoretical 
and experimental rates is provided, e.g., in the REACLIB compilation. Further 
details on REACLIB, the NON-SMOKER code, and the calculations are presented at 
\emph{\url{http://quasar.physik.unibas.ch/~tommy/reaclib.html}}.
Rates based on additional mass models can also be obtained from the authors on
request or directly at the latter URL.

\subsubsection*{\textmd{\emph{Acknowledgements}}}

This work was supported in part by the Swiss National Science Foundation (grant
2000-053798.98) and the Austrian Academy of Sciences (APART). T. R. 
acknowledges a PROFIL professorship 
of the Swiss National Science Foundation (grant 2124-055832.98).

\newcommand{\noopsort}[1]{} \newcommand{\printfirst}[2]{#1}
  \newcommand{\singleletter}[1]{#1} \newcommand{\swithchargs}[2]{#2#1}

\newpage
\section{Explanation of Tables}

\subsection*{Table I: Sample of Cross Section Tables}

Sample excerpt of the ASCII file \textit{cs\_frdm.asc}
containing cross sections for reactions
on the ground state calculated using input data
from the FRDM mass model. See RATH for details.

\begin{lyxlist}{00.00.0000}
\item [Z]Charge number of target
\item [A]Mass number of target
\item [p]Projectile: 0 for neutron, 1 for proton, 2 for \( \alpha  \) particle
\item [L]Number of calculated energies and cross sections
\item [b]Index \( b \), which assumes values of 0 or 1 depending on whether 
cross sections are listed ($b=1$) for the
first particle exit channel (defined by column f)
\item [c]Index \( c \), which assumes values of 0 or 1 depending on whether 
cross sections are listed ($c=1$) for the 
second particle exit channel (defined by column h)
\item [Zc]Charge number of the compound nucleus
\item [Ac]Mass number of the compound nucleus
\item [SN]Neutron separation energy in the compound nucleus, in MeV
\item [SP]Proton separation energy in the compound nucleus, in MeV
\item [SA]\( \alpha  \)-particle separation energy in the compound nucleus, 
in MeV
\item [T]Target written as combination of element name and mass number
\item [x]Projectile: +n for neutron, +p for proton, +a for \( \alpha  \)
particle
\item [e,~f,~h]Exit channels: \( - \)g for \( \gamma  \) ray (capture reaction),
\( - \)n for neutron, \( - \)p for proton, \( - \)a for \( \alpha  \)
particle. In this table, cross sections for the capture channel are always 
listed and column e
is always \( - \)g; in the table of reactions with \( Q<0 \) (the
\textit{cs\_frdm\_endo.asc} and \textit{cs\_etfsiq\_endo.asc} files), 
no capture reactions are given and column e is omitted.
\item [E1\ldots{}EL]Center-of-mass energies in MeV for which the nuclear 
reaction cross sections have been calculated. There are $L$ values with up to 
eight entries per line. The energies are only given once as they are the same 
for the same combination of target and projectile.
\item [G1\ldots{}GL]Capture cross sections in barn; $L$ values are given 
with up to eight entries per line. Values lower than 10\( ^{-30} \) barn are 
set equal to zero.
\item [(X1\ldots{}XL)]Nuclear cross sections for the first particle exit channel 
(specified by column f) in barn. $L$ values are given with up to eight entries 
per line. They are listed only if \( b=1 \). Values lower than 10\( ^{-30} \) 
barn are set equal to zero.
\item [(Y1\ldots{}YL)]Nuclear cross sections for the second particle channel 
(specified by column h) in barn; $L$ values are given with up to eight entries 
per line. They are listed only if \( c=1 \). Values lower than 10\( ^{-30} \) 
barn are set equal to zero.
\end{lyxlist}

\subsection*{Table II: Sample of Rate Tables}

Sample of the ASCII file \textit{rates\_frdm.asc} of laboratory rates 
\( N_{A}\left\langle \sigma v\right\rangle
^{\mathrm{lab}}  \) calculated
with the ground state cross sections of Table I
and stellar rates \( N_{A}\left\langle \sigma v\right\rangle ^* \) 
calculated from stellar cross sections with a thermally excited target,
given in cm\( ^{3} \) s\( ^{-1} \) mol\( ^{-1} \). 
Values lower than 10\( ^{-30} \) cm\( ^{3} \) s\( ^{-1} \) mol\( ^{-1} \)
are set equal to zero.

\begin{lyxlist}{00.00.0000}
\item [Z]Charge number of target
\item [A]Mass number of target
\item [p]Projectile: 0 for neutron, 1 for proton, 2 for \( \alpha  \)
particle
\item [b]Index \( b \), which assumes values of 0 or 1 depending on whether 
rates are listed ($b=1$) for the first particle exit channel (defined by 
column f)
\item [c]Index \( c \), which assumes values of 0 or 1 depending on whether 
rates are listed ($c=1$) for the second particle exit channel (defined by 
column h)
\item [T]Target written as combination of element name and mass number
\item [x]Projectile: +n for neutron, +p for proton, +a for \( \alpha  \)
particle
\item [e,~f,~h]Exit channels: \( - \)g for \( \gamma  \) ray (capture reaction),
\( - \)n for neutron, \( - \)p for proton, \( - \)a for \( \alpha  \)
particle. Rates for the capture channel are always listed and column e is 
always \( - \)g.
\item [G1\ldots{}G24]Laboratory capture rates \( N_{A}\left\langle 
\sigma v\right\rangle ^{\mathrm{lab}} \) (see Sec.~\ref{astro_exo}). 
Eight entries each in three lines are given, corresponding to the grid
of 24 temperatures
\( T_{9}= \) 0.1, 0.15, 0.2, 0.3, 0.4, 0.5, 0.6, 0.7, 0.8, 0.9, 1.0, 1.5, 2.0,
2.5, 3.0, 3.5, 4.0, 4.5, 5.0, 6.0, 7.0, 8.0, 9.0, 10.0 (\( T_{9} \) is given
in 10\( ^{9} \) K).
\item [H1\ldots{}H24]Stellar capture rates \( N_{A}\left\langle 
\sigma v\right\rangle ^* \)
(see Eq.~10 in RATH), given as eight entries each in three lines for the
24 temperature grid. Only the stellar rate should
be used in astrophysical applications and when applying detailed balance in
order to derive the reverse rate.
\item [(R1\ldots{}R24)]Laboratory rates for the first particle exit channel 
(specified by column f) at 24 temperatures. They are listed only if \( b=1 \).
\item [(U1\ldots{}U24)]Stellar rates for the first particle channel (specified 
by column f). They are listed only if \( b=1 \). Only the stellar rate should
be used in astrophysical applications and when applying detailed balance in
order to derive the reverse rate.
\item [(R1\ldots{}R24)]Laboratory rates for the second particle exit channel 
(specified by column h) at 24 temperatures. They are listed only if \( c=1 \).
\item [(U1\ldots{}U24)]Stellar rates for the second particle channel (specified 
by column h). They are listed only if \( c=1 \). Only the stellar rate should
be used in astrophysical applications and when applying detailed balance in
order to derive the reverse rate.
\end{lyxlist}

\subsection*{Table III: Sample of Nuclear Level Table}

Up to 20 experimentally known nuclear states were used in the calculation (see
Sec.~\ref{levels}). An excerpt of the full on-line ASCII file 
\textit{levels.asc} is given here as an example. Only the experimental
values used in the calculation are listed. The theoretical ground state
spins were already given in RATH.

\begin{lyxlist}{00.00.0000}
\item [Z]Charge number of the isotope
\item [A]Mass number of the isotope
\item [L]Number of given energy levels (including the ground state)
\item [T]Isotope written as a combination of element name and mass number
\item [E1\ldots{}EL]Excitation energy of the level, given in MeV
\item [p1\ldots{}pL]Parity of the level
\item [J1\ldots{}JL]Spin of the level
\end{lyxlist}


\newpage
\textheight 21.5cm
\textwidth 15cm
\singlespacing
\section*{Table I: Sample of Cross Section Tables}
\addcontentsline{toc}{section}{Table I: Sample of Cross Section Tables}
\begin{small}
\begin{verbatim}
  Z   A p  L b c     Zc  Ac   SN      SP      SA          T    x  e  f  h
   E1        E2        ...       EL
 G1        G2        ...      GL
(X1        X2        ...      XL)
(Y1        Y2        ...      YL)
--------------------------------------------------------------------------------
	.
	.
	.
 31  70 0 63 1 1     31  71   9.301   7.864   5.259      ga70 +n -g -p -a
   0.00018   0.01261   0.02565   0.03934   0.05372   0.06880   0.08463   0.10125
   0.11870   0.13701   0.15623   0.17641   0.19759   0.21982   0.24316   0.26765
   0.29336   0.32035   0.34868   0.37841   0.40962   0.44239   0.47678   0.51287
   0.55076   0.59054   0.63229   0.67611   0.72211   0.77039   0.82107   0.87427
   0.93011   0.98873   1.05025   1.11484   1.18263   1.25378   1.32847   1.40688
   1.48917   1.57555   1.66623   1.76141   1.86131   1.96618   2.19180   2.57402
   3.01606   3.52731   4.11858   4.80240   5.59326   6.50792   7.56575   8.78917
  10.20410  11.84050  13.73306  15.92186  18.45328  21.38095  24.76690
 1.096e+01 5.109e-01 3.340e-01 2.617e-01 2.202e-01 1.927e-01 1.731e-01 1.585e-01
 1.471e-01 1.379e-01 1.305e-01 1.243e-01 1.191e-01 1.146e-01 1.108e-01 1.074e-01
 1.045e-01 1.020e-01 9.980e-02 9.795e-02 9.639e-02 9.510e-02 9.408e-02 8.693e-02
 7.916e-02 7.453e-02 7.108e-02 6.592e-02 5.804e-02 5.325e-02 4.981e-02 4.724e-02
 4.042e-02 3.625e-02 3.212e-02 2.890e-02 2.627e-02 2.402e-02 2.140e-02 1.959e-02
 1.804e-02 1.629e-02 1.501e-02 1.392e-02 1.273e-02 1.181e-02 1.024e-02 8.479e-03
 7.146e-03 5.837e-03 4.513e-03 3.253e-03 2.215e-03 1.453e-03 9.323e-04 5.897e-04
 3.682e-04 2.242e-04 1.247e-04 5.079e-05 1.075e-05 1.348e-06 1.494e-07
 2.374e-02 8.829e-04 5.306e-04 4.073e-04 3.468e-04 3.134e-04 2.945e-04 2.849e-04
 2.818e-04 2.838e-04 2.902e-04 3.007e-04 3.152e-04 3.338e-04 3.568e-04 3.847e-04
 4.180e-04 4.576e-04 5.042e-04 5.592e-04 6.239e-04 7.000e-04 7.895e-04 8.593e-04
 9.425e-04 1.058e-03 1.195e-03 1.127e-03 1.180e-03 1.270e-03 1.386e-03 1.528e-03
 1.679e-03 1.834e-03 1.972e-03 2.127e-03 2.296e-03 2.477e-03 2.605e-03 2.784e-03
 2.976e-03 3.112e-03 3.294e-03 3.490e-03 3.630e-03 3.804e-03 4.132e-03 4.586e-03
 4.958e-03 5.162e-03 5.321e-03 5.535e-03 5.931e-03 6.605e-03 7.799e-03 9.642e-03
 1.221e-02 1.561e-02 2.029e-02 2.613e-02 3.209e-02 4.483e-02 2.914e-02
 3.449e-04 1.731e-05 1.217e-05 1.021e-05 9.179e-06 8.589e-06 8.255e-06 8.095e-06
 8.070e-06 8.156e-06 8.344e-06 8.628e-06 9.009e-06 9.494e-06 1.009e-05 1.081e-05
 1.168e-05 1.272e-05 1.395e-05 1.543e-05 1.718e-05 1.927e-05 2.179e-05 2.304e-05
 2.415e-05 2.634e-05 2.923e-05 3.125e-05 3.191e-05 3.461e-05 3.847e-05 4.358e-05
 4.792e-05 5.363e-05 5.948e-05 6.683e-05 7.584e-05 8.670e-05 9.715e-05 1.116e-04
 1.293e-04 1.473e-04 1.712e-04 2.008e-04 2.325e-04 2.733e-04 3.812e-04 6.464e-04
 1.117e-03 1.867e-03 2.902e-03 4.051e-03 5.188e-03 6.500e-03 8.343e-03 1.099e-02
 1.453e-02 1.912e-02 2.521e-02 3.250e-02 3.976e-02 4.676e-02 5.497e-02
 31  70 1 61 1 1     32  71   7.416   8.289   4.453      ga70 +p -g -n -a
   0.21615   0.23881   0.26240   0.28695   0.31249   0.33909   0.36676   0.39557
   0.42555   0.45675   0.48923   0.52303   0.55820   0.59482   0.63292   0.67259
   0.71386   0.75682   0.80154   0.84808   0.89651   0.94692   0.99939   1.05400
   1.11083   1.16998   1.23155   1.29562   1.36231   1.43172   1.50396   1.57915
   1.65740   1.73884   1.82361   1.91183   2.00365   2.09922   2.19868   2.30221
   2.40995   2.52208   2.88669   3.29775   3.76119   4.28367   4.87273   5.53684
   6.28557   7.12969   8.08137   9.15430  10.36395  11.72771  13.26524  14.99867
  16.95297  19.15627  21.64030  24.44082  25.45150
 9.354e-23 1.930e-21 2.927e-20 3.423e-19 3.207e-18 2.485e-17 1.599e-16 9.049e-16
 4.526e-15 2.027e-14 8.233e-14 3.064e-13 1.054e-12 3.380e-12 1.017e-11 2.889e-11
 7.788e-11 2.002e-10 4.925e-10 1.090e-09 2.363e-09 5.056e-09 1.052e-08 2.128e-08
 4.191e-08 8.056e-08 1.514e-07 2.487e-07 4.151e-07 7.153e-07 1.195e-06 1.989e-06
 3.186e-06 4.893e-06 7.600e-06 1.171e-05 1.771e-05 2.577e-05 3.738e-05 5.347e-05
 7.286e-05 9.941e-05 2.247e-04 4.087e-04 6.205e-04 8.156e-04 9.554e-04 9.983e-04
 9.314e-04 7.851e-04 6.312e-04 5.364e-04 4.253e-04 2.974e-04 1.839e-04 9.549e-05
 4.496e-05 1.745e-05 5.796e-06 1.387e-06 7.919e-07
 6.777e-22 1.498e-20 2.418e-19 3.001e-18 2.981e-17 2.447e-16 1.707e-15 1.029e-14
 5.469e-14 2.597e-13 1.116e-12 4.386e-12 1.590e-11 5.359e-11 1.690e-10 5.018e-10
 1.409e-09 3.761e-09 9.575e-09 2.341e-08 5.491e-08 1.239e-07 2.698e-07 5.685e-07
 1.161e-06 2.302e-06 4.441e-06 8.373e-06 1.537e-05 2.752e-05 4.817e-05 8.244e-05
 1.382e-04 2.271e-04 3.656e-04 5.771e-04 8.935e-04 1.358e-03 2.025e-03 2.964e-03
 4.260e-03 6.005e-03 1.503e-02 3.191e-02 5.837e-02 9.500e-02 1.421e-01 1.973e-01
 2.569e-01 3.147e-01 3.793e-01 4.956e-01 6.171e-01 6.951e-01 7.324e-01 7.529e-01
 6.897e-01 5.214e-01 3.469e-01 1.634e-01 1.151e-01
 2.036e-27 4.709e-26 8.070e-25 1.073e-23 1.150e-22 1.025e-21 7.597e-21 4.990e-20
 2.911e-19 1.528e-18 7.305e-18 3.215e-17 1.314e-16 5.026e-16 1.812e-15 6.195e-15
 2.017e-14 6.290e-14 1.885e-13 5.038e-13 1.330e-12 3.499e-12 9.008e-12 2.266e-11
 5.576e-11 1.344e-10 3.180e-10 6.749e-10 1.444e-09 3.170e-09 6.731e-09 1.434e-08
 2.971e-08 5.848e-08 1.169e-07 2.321e-07 4.542e-07 8.506e-07 1.594e-06 2.944e-06
 5.170e-06 9.104e-06 4.393e-05 1.681e-04 5.223e-04 1.348e-03 2.921e-03 5.263e-03
 8.014e-03 1.095e-02 1.486e-02 2.242e-02 3.248e-02 4.228e-02 5.106e-02 6.222e-02
 1.378e-01 2.611e-01 2.296e-01 1.002e-01 6.595e-02
 31  70 2 65 0 0     33  74   7.975   6.852   4.379      ga70 +a -g -n -p
   0.64179   0.64903   0.65743   0.66717   0.67845   0.69153   0.70668   0.72425
   0.74461   0.76821   0.79557   0.82727   0.86402   0.90662   0.95598   1.01320
   1.07952   1.15639   1.24548   1.34875   1.46844   1.60716   1.76796   1.95432
   2.17033   2.42069   2.71087   3.48925   3.55116   3.59539   3.61420   3.63436
   3.65595   3.67908   3.70385   3.73038   3.75881   3.78925   3.82186   3.85679
   3.89421   3.93429   3.97722   4.02320   4.07246   4.12522   4.18174   4.24227
   4.45097   4.70747   5.02270   5.41011   5.88625   6.47142   7.19060   8.07447
   9.16075  10.49579  12.13655  14.15305  16.63133  19.67714  23.42044  28.02096
  29.77768
 0.000e+00 0.000e+00 0.000e+00 0.000e+00 0.000e+00 0.000e+00 0.000e+00 0.000e+00
 0.000e+00 0.000e+00 0.000e+00 0.000e+00 0.000e+00 1.084e-35 2.645e-34 7.954e-33
 2.899e-31 1.255e-29 6.292e-28 3.551e-26 2.184e-24 1.417e-22 9.370e-21 6.095e-19
 3.771e-17 2.154e-15 1.102e-13 2.904e-10 4.810e-10 6.326e-10 5.406e-10 5.378e-10
 5.603e-10 5.073e-10 4.458e-10 4.625e-10 5.058e-10 5.732e-10 6.667e-10 7.851e-10
 9.441e-10 1.160e-09 1.453e-09 1.734e-09 1.611e-09 1.783e-09 2.058e-09 2.492e-09
 5.025e-09 1.269e-08 4.171e-08 1.707e-07 7.990e-07 3.932e-06 1.861e-05 7.506e-05
 2.181e-04 3.652e-04 3.597e-04 2.483e-04 1.177e-04 2.259e-05 1.377e-06 4.875e-08
 1.987e-08
 31  71 0 60 0 1     31  72   6.521   8.551   5.467      ga71 +n -g -p -a
   0.00018   0.00423   0.00900   0.01463   0.02125   0.02907   0.03828   0.04914
   0.06193   0.07702   0.09480   0.11576   0.14047   0.16960   0.20393   0.24440
   0.29211   0.34834   0.41463   0.49277   0.58487   0.69345   0.82143   0.97229
   1.15011   1.35973   1.60682   1.89809   2.24142   2.27455   2.30962   2.34677
   2.38610   2.42775   2.47186   2.51857   2.56803   2.62041   2.67588   2.73461
   2.79681   2.86268   3.08450   3.34790   3.66070   4.03213   4.47320   4.99696
   5.61892   6.35749   7.23452   8.27598   9.51269  10.98127  12.72517  14.79603
  17.25513  20.17527  23.64288  27.76060
 4.126e+00 3.253e-01 1.957e-01 1.436e-01 1.135e-01 9.312e-02 7.829e-02 6.698e-02
 5.811e-02 5.102e-02 4.524e-02 4.048e-02 3.651e-02 3.321e-02 3.045e-02 2.817e-02
 2.632e-02 2.486e-02 2.316e-02 1.990e-02 1.483e-02 1.243e-02 1.120e-02 1.033e-02
 8.538e-03 7.404e-03 5.946e-03 5.439e-03 5.251e-03 5.225e-03 5.169e-03 5.132e-03
 5.093e-03 5.032e-03 4.976e-03 4.923e-03 4.871e-03 4.783e-03 4.714e-03 4.647e-03
 4.552e-03 4.461e-03 4.157e-03 3.789e-03 3.351e-03 2.848e-03 2.310e-03 1.757e-03
 1.264e-03 8.680e-04 5.741e-04 3.699e-04 3.043e-04 3.016e-04 4.249e-04 5.691e-04
 2.808e-04 3.937e-05 3.016e-06 1.434e-07
 3.111e-26 2.868e-27 2.098e-27 1.964e-27 2.080e-27 2.417e-27 3.058e-27 4.221e-27
 6.396e-27 1.073e-26 2.016e-26 4.292e-26 1.049e-25 2.980e-25 9.974e-25 3.981e-24
 1.915e-23 1.119e-22 7.479e-22 5.784e-21 4.563e-20 4.925e-19 6.664e-18 1.025e-16
 1.666e-15 3.099e-14 5.916e-13 1.107e-11 1.990e-10 2.560e-10 3.313e-10 4.343e-10
 5.753e-10 7.680e-10 1.037e-09 1.415e-09 1.953e-09 2.709e-09 3.809e-09 5.412e-09
 7.733e-09 1.117e-08 3.540e-08 1.195e-07 4.216e-07 1.517e-06 5.442e-06 1.875e-05
 6.141e-05 1.874e-04 5.147e-04 1.214e-03 3.119e-03 8.642e-03 3.286e-02 1.408e-01
 4.250e-01 6.735e-01 7.452e-01 9.032e-01
	.
	.
	.
 32  55 0 63 1 1     32  56  22.189  -1.385   4.248      ge55 +n -g -p -a
   0.00018   0.00662   0.01337   0.02046   0.02789   0.03569   0.04387   0.05245
   0.06145   0.07090   0.08081   0.09121   0.10211   0.11356   0.12557   0.13816
   0.15138   0.16524   0.17979   0.19505   0.21106   0.22786   0.24549   0.26398
   0.28337   0.30373   0.32508   0.34748   0.37098   0.39564   0.42150   0.44864
   0.47712   0.50699   0.53833   0.57121   0.60570   0.64189   0.67986   0.71970
   0.76149   0.80534   0.85134   0.89960   0.95023   1.11755   1.31077   1.53390
   1.79156   2.08911   2.43271   2.82949   3.28769   3.81681   4.42782   5.13341
   5.94822   6.88914   7.97569   9.23043  10.67938  12.35259  12.96602
 6.622e-04 1.076e-04 7.536e-05 6.094e-05 5.237e-05 4.655e-05 4.229e-05 3.904e-05
 3.644e-05 3.431e-05 3.255e-05 3.106e-05 2.978e-05 2.869e-05 2.774e-05 2.692e-05
 2.619e-05 2.556e-05 2.500e-05 2.451e-05 2.408e-05 2.371e-05 2.338e-05 2.319e-05
 2.294e-05 2.273e-05 2.254e-05 2.238e-05 2.225e-05 2.253e-05 2.243e-05 2.235e-05
 2.228e-05 2.223e-05 2.220e-05 2.217e-05 2.215e-05 2.214e-05 2.213e-05 2.213e-05
 2.213e-05 2.212e-05 2.211e-05 2.237e-05 2.234e-05 2.217e-05 2.188e-05 2.152e-05
 2.105e-05 2.059e-05 2.073e-05 2.110e-05 2.302e-05 2.547e-05 2.917e-05 3.154e-05
 3.022e-05 2.724e-05 2.548e-05 2.534e-05 2.481e-05 2.285e-05 2.177e-05
 6.033e+01 9.802e+00 6.859e+00 5.544e+00 4.761e+00 4.229e+00 3.840e+00 3.540e+00
 3.301e+00 3.106e+00 2.944e+00 2.807e+00 2.689e+00 2.588e+00 2.499e+00 2.422e+00
 2.354e+00 2.294e+00 2.241e+00 2.194e+00 2.152e+00 2.115e+00 2.081e+00 2.061e+00
 2.035e+00 2.011e+00 1.990e+00 1.972e+00 1.955e+00 1.940e+00 1.927e+00 1.916e+00
 1.905e+00 1.896e+00 1.887e+00 1.880e+00 1.873e+00 1.866e+00 1.860e+00 1.854e+00
 1.848e+00 1.841e+00 1.835e+00 1.828e+00 1.821e+00 1.798e+00 1.770e+00 1.738e+00
 1.705e+00 1.678e+00 1.661e+00 1.660e+00 1.676e+00 1.699e+00 1.714e+00 1.708e+00
 1.681e+00 1.646e+00 1.614e+00 1.588e+00 1.560e+00 1.531e+00 1.523e+00
 3.348e-02 5.449e-03 3.819e-03 3.093e-03 2.661e-03 2.369e-03 2.156e-03 1.993e-03
 1.863e-03 1.758e-03 1.671e-03 1.597e-03 1.535e-03 1.482e-03 1.436e-03 1.397e-03
 1.362e-03 1.333e-03 1.308e-03 1.286e-03 1.267e-03 1.251e-03 1.238e-03 1.232e-03
 1.223e-03 1.216e-03 1.211e-03 1.208e-03 1.206e-03 1.226e-03 1.226e-03 1.228e-03
 1.230e-03 1.234e-03 1.239e-03 1.244e-03 1.250e-03 1.256e-03 1.263e-03 1.271e-03
 1.279e-03 1.286e-03 1.294e-03 1.318e-03 1.324e-03 1.339e-03 1.346e-03 1.347e-03
 1.340e-03 1.337e-03 1.386e-03 1.478e-03 1.722e-03 2.050e-03 2.496e-03 2.836e-03
 2.953e-03 3.013e-03 3.060e-03 3.382e-03 4.254e-03 5.206e-03 5.496e-03
	.
	.
	.
 32  70 0 66 0 1     32  71   7.416   8.289   4.453      ge70 +n -g -p -a
   0.00018   0.00385   0.00875   0.01528   0.02398   0.03559   0.05106   0.07170
   0.09923   0.13593   0.18488   0.25014   0.33718   0.45325   0.60803   0.81442
   1.08966   1.11732   1.14635   1.17683   1.20883   1.24242   1.27769   1.31470
   1.35356   1.39436   1.43718   1.48214   1.52934   1.57888   1.63089   1.68549
   1.74281   1.80298   1.86614   1.93245   2.00207   2.07514   2.15185   2.23238
   2.31692   2.40567   2.49884   2.59664   2.69931   2.80710   2.92024   3.03902
   3.16371   3.29461   3.43203   3.88669   4.41268   5.02120   5.72518   6.53961
   7.48182   8.57184   9.83288  11.29176  12.97951  14.93206  17.19093  19.80419
  22.82745  26.32501
 3.742e+00 3.301e-01 2.001e-01 1.466e-01 1.156e-01 9.436e-02 7.872e-02 6.697e-02
 5.820e-02 5.162e-02 4.679e-02 4.362e-02 4.206e-02 4.200e-02 4.349e-02 4.716e-02
 3.867e-02 3.662e-02 3.500e-02 3.369e-02 3.262e-02 3.098e-02 2.999e-02 2.922e-02
 2.863e-02 2.820e-02 2.791e-02 2.775e-02 2.772e-02 2.782e-02 2.804e-02 2.838e-02
 2.712e-02 2.651e-02 2.622e-02 2.617e-02 2.634e-02 2.671e-02 2.729e-02 2.541e-02
 2.464e-02 2.428e-02 2.383e-02 2.269e-02 2.198e-02 2.171e-02 2.143e-02 2.110e-02
 2.065e-02 1.999e-02 1.916e-02 1.685e-02 1.404e-02 1.087e-02 7.744e-03 5.092e-03
 3.088e-03 1.780e-03 9.942e-04 5.430e-04 2.828e-04 1.216e-04 3.075e-05 4.647e-06
 5.384e-07 1.075e-07
 2.169e-08 2.164e-09 1.445e-09 1.160e-09 1.012e-09 9.343e-10 9.106e-10 9.480e-10
 1.072e-09 1.339e-09 1.890e-09 3.109e-09 6.170e-09 1.527e-08 4.862e-08 2.062e-07
 8.423e-07 9.275e-07 1.037e-06 1.175e-06 1.347e-06 1.518e-06 1.757e-06 2.059e-06
 2.440e-06 2.924e-06 3.541e-06 4.335e-06 5.363e-06 6.703e-06 8.465e-06 1.080e-05
 1.302e-05 1.625e-05 2.066e-05 2.665e-05 3.485e-05 4.615e-05 6.188e-05 7.816e-05
 1.021e-04 1.352e-04 1.792e-04 2.318e-04 3.055e-04 4.107e-04 5.548e-04 7.465e-04
 1.003e-03 1.333e-03 1.759e-03 4.031e-03 8.628e-03 1.655e-02 2.724e-02 3.746e-02
 4.420e-02 5.027e-02 5.884e-02 7.042e-02 8.494e-02 1.133e-01 1.958e-01 1.222e-01
 3.338e-02 1.103e-02
 32  70 1 79 0 1     33  71  11.623   4.621   3.440      ge70 +p -g -n -a
   0.22119   0.23140   0.24232   0.25398   0.26643   0.27974   0.29396   0.30915
   0.32538   0.34272   0.36125   0.38105   0.40219   0.42479   0.44893   0.47472
   0.50228   0.53172   0.56318   0.59678   0.63269   0.67105   0.71204   0.75583
   0.80261   0.85260   0.90601   0.96306   1.02402   1.08916   1.15874   1.23309
   1.31252   1.39739   1.48806   1.58493   1.68843   1.79901   1.91715   2.04338
   2.17824   2.32232   2.47626   2.64073   2.81645   3.00419   3.20478   3.41908
   3.64804   3.89267   4.15403   4.43327   4.73160   5.05035   5.39090   5.75474
   6.14348   6.55880   6.94929   6.98035   7.00253   7.03142   7.06414   7.10119
   7.14315   7.19068   7.24450   7.30546   7.37450   7.64153   8.02943   8.59292
   9.41147  10.60054  12.32785  14.83703  18.48201  23.77690  28.57977
 2.284e-22 9.721e-22 4.119e-21 1.735e-20 7.245e-20 2.996e-19 1.224e-18 4.936e-18
 1.960e-17 7.659e-17 2.940e-16 1.107e-15 4.085e-15 1.476e-14 5.216e-14 1.802e-13
 6.079e-13 2.002e-12 6.433e-12 2.016e-11 6.159e-11 1.834e-10 5.321e-10 1.504e-09
 4.144e-09 1.112e-08 2.908e-08 7.407e-08 1.837e-07 4.433e-07 1.039e-06 2.357e-06
 5.141e-06 1.068e-05 2.092e-05 3.855e-05 6.736e-05 1.130e-04 1.829e-04 2.845e-04
 4.241e-04 6.078e-04 8.439e-04 1.139e-03 1.498e-03 1.920e-03 2.400e-03 2.935e-03
 3.517e-03 4.141e-03 4.805e-03 5.486e-03 6.157e-03 6.758e-03 7.206e-03 7.445e-03
 7.363e-03 6.962e-03 6.532e-03 6.510e-03 6.495e-03 6.455e-03 6.395e-03 6.324e-03
 5.977e-03 5.892e-03 5.776e-03 5.632e-03 5.463e-03 4.647e-03 3.758e-03 2.890e-03
 2.060e-03 1.224e-03 5.621e-04 1.944e-04 2.597e-05 9.259e-07 2.936e-08
 7.400e-43 4.444e-42 2.709e-41 1.675e-40 1.048e-39 6.640e-39 4.250e-38 2.747e-37
 1.791e-36 1.178e-35 7.802e-35 5.204e-34 3.494e-33 2.359e-32 1.599e-31 1.091e-30
 7.480e-30 5.148e-29 3.557e-28 2.465e-27 1.713e-26 1.193e-25 8.322e-25 5.810e-24
 4.057e-23 2.830e-22 1.970e-21 1.367e-20 9.441e-20 6.471e-19 4.383e-18 2.924e-17
 1.902e-16 1.191e-15 7.093e-15 3.991e-14 2.136e-13 1.099e-12 5.458e-12 2.598e-11
 1.179e-10 5.101e-10 2.111e-09 8.380e-09 3.189e-08 1.159e-07 4.025e-07 1.332e-06
 4.199e-06 1.262e-05 3.620e-05 9.919e-05 2.593e-04 6.429e-04 1.497e-03 3.253e-03
 6.483e-03 1.178e-02 1.883e-02 1.952e-02 2.002e-02 2.061e-02 2.123e-02 2.195e-02
 2.181e-02 2.272e-02 2.368e-02 2.473e-02 2.589e-02 2.914e-02 3.388e-02 4.108e-02
 5.266e-02 7.042e-02 9.488e-02 1.221e-01 2.332e-01 1.433e-01 9.894e-03
 32  70 2 75 0 0     34  74  12.068   8.545   4.077      ge70 +a -g -n -p
   0.65552   0.66404   0.67339   0.68362   0.69484   0.70713   0.72059   0.73534
   0.75150   0.76921   0.78862   0.80988   0.83317   0.85870   0.88666   0.91731
   0.95088   0.98767   1.02797   1.07213   1.12052   1.17353   1.23162   1.29527
   1.36500   1.44141   1.52513   1.61686   1.71736   1.82748   1.94814   2.08034
   2.22519   2.38389   2.55779   2.74831   2.95707   3.18581   3.43642   3.71102
   4.01188   4.34154   4.70273   4.76759   4.89054   5.12364   5.56552   6.40321
   7.80068   7.91185   7.99125   8.01439   8.03991   8.06805   8.09910   8.13334
   8.17111   8.21276   8.25870   8.30937   8.36526   8.42690   8.49488   8.74379
   9.07774   9.52582  10.12700  10.93362  12.01588  13.46795  15.41621  18.03023
  21.53748  26.24321  30.24312
 0.000e+00 0.000e+00 0.000e+00 0.000e+00 0.000e+00 0.000e+00 0.000e+00 0.000e+00
 0.000e+00 0.000e+00 0.000e+00 0.000e+00 0.000e+00 0.000e+00 0.000e+00 0.000e+00
 0.000e+00 8.548e-35 9.326e-34 1.091e-32 1.361e-31 1.798e-30 2.496e-29 3.614e-28
 5.414e-27 8.311e-26 1.299e-24 2.048e-23 3.227e-22 5.039e-21 7.736e-20 1.157e-18
 1.676e-17 2.332e-16 3.097e-15 3.900e-14 4.624e-13 5.144e-12 5.338e-11 5.139e-10
 4.578e-09 3.754e-08 2.819e-07 3.937e-07 7.262e-07 2.155e-06 1.336e-05 1.943e-04
 2.062e-03 1.878e-03 2.004e-03 1.920e-03 1.776e-03 1.652e-03 1.539e-03 1.451e-03
 1.326e-03 1.223e-03 1.150e-03 1.073e-03 1.024e-03 9.863e-04 9.686e-04 9.189e-04
 9.455e-04 1.122e-03 1.364e-03 1.488e-03 1.296e-03 8.147e-04 3.616e-04 1.082e-04
 1.096e-05 2.904e-07 1.185e-08
	.
	.
	.
\end{verbatim}
\end{small}
\newpage
\section*{Table II: Sample of Rate Tables}
\addcontentsline{toc}{section}{Table II: Sample of Rate Tables}
\begin{small}
\singlespacing
\begin{verbatim}
  Z   A p b c       T    x  e  f  h
  G1        G2        ...G24

  H1        H2        ...H24

( R1        R2        ...R24 )
( U1        U2        ...U24 )
--------------------------------------------------------------------------------
	.
	.
	.
 31  70 0 1 1      ga70 +n -g -p -a
  4.64e+07  4.54e+07  4.50e+07  4.44e+07  4.41e+07  4.39e+07  4.39e+07  4.39e+07
  4.40e+07  4.42e+07  4.43e+07  4.52e+07  4.56e+07  4.54e+07  4.48e+07  4.40e+07
  4.29e+07  4.18e+07  4.06e+07  3.82e+07  3.60e+07  3.40e+07  3.21e+07  3.04e+07

  4.64e+07  4.54e+07  4.50e+07  4.44e+07  4.41e+07  4.39e+07  4.39e+07  4.39e+07
  4.40e+07  4.41e+07  4.43e+07  4.45e+07  4.29e+07  3.91e+07  3.43e+07  2.94e+07
  2.49e+07  2.10e+07  1.77e+07  1.25e+07  8.79e+06  6.12e+06  4.22e+06  2.89e+06

  8.53e+04  7.95e+04  7.69e+04  7.50e+04  7.55e+04  7.77e+04  8.12e+04  8.59e+04
  9.18e+04  9.88e+04  1.07e+05  1.62e+05  2.39e+05  3.32e+05  4.41e+05  5.62e+05
  6.94e+05  8.34e+05  9.80e+05  1.29e+06  1.60e+06  1.92e+06  2.24e+06  2.56e+06

  8.53e+04  7.95e+04  7.69e+04  7.50e+04  7.55e+04  7.77e+04  8.12e+04  8.60e+04
  9.20e+04  9.93e+04  1.08e+05  1.74e+05  2.77e+05  4.09e+05  5.54e+05  7.06e+05
  8.60e+05  1.02e+06  1.18e+06  1.54e+06  2.01e+06  2.69e+06  3.71e+06  5.16e+06

  1.57e+03  1.59e+03  1.63e+03  1.73e+03  1.83e+03  1.95e+03  2.10e+03  2.26e+03
  2.45e+03  2.66e+03  2.90e+03  4.45e+03  6.66e+03  9.75e+03  1.40e+04  2.00e+04
  2.81e+04  3.89e+04  5.32e+04  9.40e+04  1.55e+05  2.38e+05  3.45e+05  4.78e+05

  1.57e+03  1.59e+03  1.63e+03  1.73e+03  1.83e+03  1.95e+03  2.10e+03  2.27e+03
  2.46e+03  2.70e+03  2.97e+03  5.34e+03  1.07e+04  2.14e+04  4.06e+04  7.28e+04
  1.24e+05  2.00e+05  3.11e+05  6.78e+05  1.33e+06  2.39e+06  4.01e+06  6.33e+06
 31  70 1 1 1      ga70 +p -g -n -a
  4.69e-23  2.93e-18  2.77e-15  1.43e-11  3.07e-09  1.36e-07  2.38e-06  2.32e-05
  1.50e-04  7.17e-04  2.74e-03  2.93e-01  5.21e+00  3.81e+01  1.65e+02  5.10e+02
  1.25e+03  2.58e+03  4.72e+03  1.21e+04  2.45e+04  4.22e+04  6.50e+04  9.24e+04

  4.69e-23  2.93e-18  2.77e-15  1.43e-11  3.07e-09  1.36e-07  2.38e-06  2.32e-05
  1.50e-04  7.17e-04  2.74e-03  2.94e-01  5.30e+00  3.85e+01  1.60e+02  4.62e+02
  1.04e+03  1.97e+03  3.27e+03  6.85e+03  1.10e+04  1.48e+04  1.73e+04  1.82e+04

  3.94e-22  2.96e-17  3.27e-14  2.11e-10  5.27e-08  2.65e-06  5.22e-05  5.61e-04
  3.96e-03  2.06e-02  8.46e-02  1.18e+01  2.48e+02  2.02e+03  9.56e+03  3.18e+04
  8.32e+04  1.83e+05  3.56e+05  1.02e+06  2.28e+06  4.32e+06  7.27e+06  1.12e+07

  3.94e-22  2.96e-17  3.27e-14  2.11e-10  5.27e-08  2.65e-06  5.22e-05  5.61e-04
  3.96e-03  2.06e-02  8.46e-02  1.18e+01  2.48e+02  2.02e+03  9.55e+03  3.17e+04
  8.30e+04  1.83e+05  3.53e+05  1.01e+06  2.23e+06  4.18e+06  6.94e+06  1.05e+07

  1.37e-27  1.28e-22  1.80e-19  1.87e-15  7.25e-13  5.36e-11  1.51e-09  2.26e-08
  2.18e-07  1.52e-06  8.23e-06  3.68e-03  1.79e-01  2.72e+00  2.05e+01  9.86e+01
  3.47e+02  9.74e+02  2.31e+03  9.11e+03  2.58e+04  5.88e+04  1.15e+05  2.01e+05

  1.37e-27  1.28e-22  1.80e-19  1.87e-15  7.25e-13  5.36e-11  1.51e-09  2.27e-08
  2.19e-07  1.53e-06  8.36e-06  4.11e-03  2.34e-01  4.13e+00  3.51e+01  1.83e+02
  6.77e+02  1.97e+03  4.80e+03  1.98e+04  5.89e+04  1.43e+05  3.09e+05  6.21e+05
 31  70 2 0 0      ga70 +a -g -n -p
  6.12e-71  1.05e-55  1.25e-47  4.71e-38  4.65e-32  8.29e-28  1.45e-24  5.67e-22
  7.30e-20  3.96e-18  1.15e-16  1.42e-11  9.20e-09  5.47e-07  1.14e-05  1.44e-04
  1.26e-03  7.92e-03  3.79e-02  4.57e-01  2.98e+00  1.28e+01  4.11e+01  1.06e+02

  6.12e-71  1.05e-55  1.25e-47  4.71e-38  4.65e-32  8.29e-28  1.45e-24  5.67e-22
  7.30e-20  3.95e-18  1.15e-16  1.37e-11  8.03e-09  4.24e-07  8.58e-06  1.10e-04
  9.49e-04  5.68e-03  2.51e-02  2.49e-01  1.31e+00  4.43e+00  1.10e+01  2.18e+01
 31  71 0 0 1      ga71 +n -g -p -a
  1.62e+07  1.52e+07  1.46e+07  1.38e+07  1.34e+07  1.30e+07  1.28e+07  1.26e+07
  1.25e+07  1.24e+07  1.23e+07  1.21e+07  1.18e+07  1.16e+07  1.13e+07  1.10e+07
  1.07e+07  1.05e+07  1.02e+07  9.81e+06  9.44e+06  9.12e+06  8.83e+06  8.56e+06

  1.62e+07  1.52e+07  1.46e+07  1.38e+07  1.34e+07  1.30e+07  1.28e+07  1.26e+07
  1.25e+07  1.24e+07  1.23e+07  1.16e+07  1.09e+07  1.01e+07  9.39e+06  8.76e+06
  8.17e+06  7.58e+06  6.96e+06  5.62e+06  4.23e+06  2.99e+06  2.02e+06  1.33e+06

  2.23e-19  3.42e-19  6.55e-19  5.95e-18  1.13e-16  2.20e-15  3.69e-14  4.87e-13
  4.98e-12  4.01e-11  2.62e-10  3.35e-07  3.87e-05  1.16e-03  1.53e-02  1.17e-01
  6.14e-01  2.44e+00  7.87e+00  5.22e+01  2.29e+02  7.62e+02  2.11e+03  5.13e+03

  2.23e-19  3.42e-19  6.55e-19  6.14e-18  1.52e-16  4.34e-15  9.83e-14  1.64e-12
  2.04e-11  1.95e-10  1.50e-09  3.65e-06  7.14e-04  3.51e-02  7.35e-01  8.58e+00
  6.53e+01  3.57e+02  1.50e+03  1.44e+04  7.55e+04  2.56e+05  6.32e+05  1.23e+06
	.
	.
	.
 32  55 0 1 1      ge55 +n -g -p -a
  6.78e+03  7.05e+03  7.20e+03  7.39e+03  7.54e+03  7.68e+03  7.83e+03  7.98e+03
  8.13e+03  8.28e+03  8.44e+03  9.24e+03  1.00e+04  1.08e+04  1.16e+04  1.23e+04
  1.30e+04  1.36e+04  1.42e+04  1.54e+04  1.66e+04  1.77e+04  1.88e+04  1.99e+04

  7.31e+03  7.87e+03  8.28e+03  8.85e+03  9.23e+03  9.48e+03  9.65e+03  9.78e+03
  9.89e+03  9.98e+03  1.01e+04  1.05e+04  1.10e+04  1.15e+04  1.20e+04  1.25e+04
  1.30e+04  1.35e+04  1.40e+04  1.52e+04  1.66e+04  1.80e+04  1.93e+04  2.04e+04

  6.17e+08  6.41e+08  6.54e+08  6.71e+08  6.83e+08  6.96e+08  7.08e+08  7.20e+08
  7.33e+08  7.45e+08  7.58e+08  8.22e+08  8.84e+08  9.44e+08  1.00e+09  1.05e+09
  1.11e+09  1.15e+09  1.20e+09  1.29e+09  1.37e+09  1.44e+09  1.51e+09  1.58e+09

  6.70e+08  7.23e+08  7.63e+08  8.20e+08  8.58e+08  8.83e+08  9.02e+08  9.15e+08
  9.26e+08  9.35e+08  9.43e+08  9.80e+08  1.02e+09  1.05e+09  1.09e+09  1.13e+09
  1.17e+09  1.20e+09  1.24e+09  1.30e+09  1.36e+09  1.42e+09  1.47e+09  1.51e+09

  3.44e+05  3.58e+05  3.66e+05  3.76e+05  3.85e+05  3.93e+05  4.02e+05  4.11e+05
  4.20e+05  4.29e+05  4.39e+05  4.88e+05  5.38e+05  5.88e+05  6.37e+05  6.85e+05
  7.32e+05  7.77e+05  8.21e+05  9.08e+05  9.96e+05  1.09e+06  1.18e+06  1.28e+06

  3.70e+05  3.99e+05  4.20e+05  4.51e+05  4.71e+05  4.85e+05  4.96e+05  5.05e+05
  5.12e+05  5.19e+05  5.26e+05  5.63e+05  6.03e+05  6.46e+05  6.90e+05  7.35e+05
	.
	.
	.
 32  70 0 0 1      ge70 +n -g -p -a
  1.66e+07  1.61e+07  1.58e+07  1.57e+07  1.56e+07  1.57e+07  1.58e+07  1.60e+07
  1.62e+07  1.64e+07  1.66e+07  1.80e+07  1.94e+07  2.08e+07  2.20e+07  2.30e+07
  2.39e+07  2.46e+07  2.52e+07  2.62e+07  2.69e+07  2.75e+07  2.78e+07  2.81e+07

  1.66e+07  1.61e+07  1.58e+07  1.57e+07  1.56e+07  1.57e+07  1.58e+07  1.60e+07
  1.62e+07  1.64e+07  1.66e+07  1.80e+07  1.96e+07  2.13e+07  2.29e+07  2.42e+07
  2.53e+07  2.59e+07  2.59e+07  2.40e+07  2.01e+07  1.52e+07  1.06e+07  6.90e+06

  1.24e-01  1.32e-01  1.41e-01  1.67e-01  2.01e-01  2.49e-01  3.16e-01  4.11e-01
  5.51e-01  7.54e-01  1.05e+00  5.77e+00  2.88e+01  1.24e+02  4.55e+02  1.41e+03
  3.73e+03  8.65e+03  1.79e+04  5.93e+04  1.50e+05  3.17e+05  5.82e+05  9.64e+05

  1.24e-01  1.32e-01  1.41e-01  1.67e-01  2.01e-01  2.49e-01  3.16e-01  4.12e-01
  5.52e-01  7.63e-01  1.09e+00  9.52e+00  9.50e+01  7.32e+02  4.23e+03  1.88e+04
  6.72e+04  1.98e+05  4.97e+05  2.14e+06  6.22e+06  1.38e+07  2.52e+07  4.01e+07
 32  70 1 0 1      ge70 +p -g -n -a
  7.61e-23  7.14e-18  9.14e-15  7.11e-11  2.01e-08  1.10e-06  2.32e-05  2.61e-04
  1.90e-03  9.95e-03  4.07e-02  4.80e+00  7.80e+01  5.00e+02  1.92e+03  5.34e+03
  1.20e+04  2.33e+04  4.04e+04  9.64e+04  1.86e+05  3.11e+05  4.70e+05  6.59e+05

  7.61e-23  7.14e-18  9.14e-15  7.11e-11  2.01e-08  1.10e-06  2.32e-05  2.61e-04
  1.90e-03  9.95e-03  4.07e-02  4.80e+00  7.79e+01  5.01e+02  1.92e+03  5.35e+03
  1.20e+04  2.28e+04  3.84e+04  8.10e+04  1.25e+05  1.51e+05  1.50e+05  1.29e+05

  3.34e-42  7.00e-36  1.36e-31  9.30e-26  8.93e-22  8.02e-19  1.56e-16  1.08e-14
  3.60e-13  7.02e-12  9.11e-11  7.75e-07  2.23e-04  1.19e-02  2.32e-01  2.37e+00
  1.53e+01  7.05e+01  2.53e+02  1.88e+03  8.45e+03  2.72e+04  6.93e+04  1.49e+05

  3.34e-42  7.00e-36  1.36e-31  9.30e-26  8.93e-22  8.02e-19  1.56e-16  1.08e-14
  3.62e-13  7.11e-12  9.39e-11  1.02e-06  4.28e-04  3.26e-02  8.75e-01  1.17e+01
  9.34e+01  5.11e+02  2.07e+03  1.77e+04  8.19e+04  2.55e+05  6.11e+05  1.21e+06
 32  70 2 0 0      ge70 +a -g -n -p
  3.70e-74  1.33e-57  5.31e-49  3.86e-39  5.15e-33  1.13e-28  2.26e-25  9.58e-23
  1.38e-20  9.11e-19  3.31e-17  9.48e-12  2.29e-08  5.01e-06  2.56e-04  5.03e-03
  5.12e-02  3.28e-01  1.50e+00  1.57e+01  8.73e+01  3.23e+02  9.01e+02  2.05e+03

  3.70e-74  1.33e-57  5.31e-49  3.86e-39  5.15e-33  1.13e-28  2.26e-25  9.58e-23
  1.38e-20  9.11e-19  3.31e-17  9.48e-12  2.29e-08  4.98e-06  2.48e-04  4.64e-03
  4.40e-02  2.56e-01  1.04e+00  7.79e+00  2.81e+01  6.19e+01  9.75e+01  1.22e+02
	.
	.
	.
\end{verbatim}
\end{small}
\newpage
\section*{Table III: Sample of Nuclear Level Table}
\addcontentsline{toc}{section}{Table III: Sample of Nuclear Level Table}
\begin{small}
\singlespacing
\begin{verbatim}
 Z  A L  T
 E1     p1 J1  E2     p2 J2 ...            EL     pL JL
--------------------------------------------------------------------------------
	.
	.
	.
31 69 6 ga69
 0.0000  -1.5  0.3187  -0.5  0.5742  -2.5  0.8721  -1.5  1.0286  -0.5
 1.1070  -2.5 
31 70 5 ga70
 0.0000  +1.0  0.5082  +2.0  0.6511  +1.0  0.6909  -2.0  0.8791  -4.0
31 71 20 ga71
 0.0000  -1.5  0.3900  -0.5  0.4873  -2.5  0.5115  -1.5  0.9101  -1.5
 0.9647  -2.5  1.1074  -3.5  1.1093  -0.5  1.3950  -3.5  1.4759  -2.5
 1.4937  +4.5  1.4984  +2.5  1.6312  -1.5  1.7022  +0.5  1.7197  -2.5
 1.7524  -1.5  1.9052  -2.5  1.9416  -1.5  1.9950  -2.5  2.0646  -0.5
	.
	.
	.
32 61 1 ge61
 0.0000  -1.5
32 64 11 ge64
 0.0000  +0.0  0.9020  +2.0  1.5780  +2.0  2.0530  +4.0  2.9690  -3.0
 3.4670  +6.0  3.7170  -5.0  4.2440  -7.0  5.1810  +8.0  5.3710  -9.0
 6.5630 -11.0
	.
	.
	.
32 70 20 ge70
 0.0000  +0.0  1.0393  +2.0  1.2154  +0.0  1.7079  +2.0  2.1535  +4.0
 2.1574  +2.0  2.3069  +0.0  2.4515  +3.0  2.5357  +2.0  2.5614  -3.0
 2.8067  +4.0  2.8878  +0.0  2.9452  +2.0  3.0468  +3.0  3.0589  +4.0
 3.1070  +0.0  3.1804  +2.0  3.1942  +4.0  3.2420  +1.0  3.2961  +3.0
32 71 20 ge71
 0.0000  -0.5  0.1749  -2.5  0.1984  +4.5  0.4999  -1.5  0.5251  +2.5
 0.5898  +3.5  0.7082  -1.5  0.7473  -2.5  0.8083  -0.5  0.8313  -1.5
 0.8869  -1.5  1.0266  -2.5  1.0382  +4.5  1.0955  -1.5  1.0961  +3.5
 1.1394  +1.5  1.1710  +2.5  1.1724  +6.5  1.1923  +5.5  1.2051  +2.5
	.
	.
	.
\end{verbatim}
\end{small}

\end{document}